\documentclass[a4paper,12pt]{article}
\usepackage[russian]{babel}
\usepackage{amsmath}
\usepackage{ulem}
\usepackage{calc}
\usepackage{vmargin}
\usepackage{amstext}
\usepackage{graphicx}
\usepackage{amsfonts}
\usepackage{amssymb}

\begin{document}
	%	\author{О.В. Капцов\\ Институт вычислительного моделирования, Россия \\ email: kaptsov@icm.krasn.ru \\ \\
	%		Д.О. Капцов\\ Институт вычислительного моделирования, Россия \\ email: hot.dok@gmail.com }
	%\maketitle	
	
	\title{%\raggedright УДК 532.5+517.95  \\
		{\bf Integration  of acoustic wave equations  for inhomogeneous media.   }}
	
%Integration of equation for a acoustic  propagation in inhomogeneous   media} 
%}
	
	\author{ \bf O.V. Kaptsov
		\\ Institute of Computational Modelling, SD of RAS  
		\\ 660036, Krasnoyarsk, Russia
		\\ Email: kaptsov@icm.krasn.ru}

	\date{}
	\maketitle
	
	We obtain exact solutions of the acoustic wave equations for inhomogeneous media.
	Two methods for integrating these equations are proposed. The first one is based on the Laplace cascade method, while the second method involves reducing two-dimensional and three-dimensional models to the wave equation. In the case of plane waves, we find new solutions  depending on two arbitrary functions. These solutions generalize the classical ones obtained by Euler. In the two-dimensional and three-dimensional cases, equations that can be reduced to equations with constant coefficients are found.
	

\numberwithin{equation}{section}

\section{Introduction}

Real media are inhomogeneous, and therefore studies on wave propagation in such media are of great interest and have a wide range of applications \cite{Br, Ostash, Mei,Johnson}. A classical object of study is acoustic waves, which are typically described by the linear equation 
\begin{equation} \label{Ptt} p_{tt} = c^2 \Delta p , \end{equation}
 where $p$ is the sound pressure, $\Delta$ is the Laplace operator, and $c$ is the speed of sound, which may depend on the coordinates
$ x,y,z$.  
Leonhard Euler was the first to obtain general solutions \cite{Euler} of the equation
\begin{equation} \label{p} 
	p_{tt} = c^2 p_{xx} , 
\end{equation} 
with a non-constant coefficient
$$ c^2 = x^{\frac{4m}{2m-1}}  ,$$
where $m$ is an arbitrary integer. 
In what follows, a solution of the equation (\ref{p}) will be called general if it depends on two arbitrary functions.
Recently, such solutions have found applications in shallow water theory and acoustics \cite{Pel1,Pel2}. For other coefficients,  as far as we know, general solutions of equation (\ref{p}) have not been obtained. 
As for the multidimensional equation (\ref{Ptt}), integral representations for its solutions in the case $c^2=1$ are well known.
 On the other hand, Smirnov and Sobolev \cite{Sobolev}, as well as Erugin \cite{Erugin}, obtained functionally invariant solutions to the wave equation (\ref{Ptt}) with a constant coefficient $c^2$
in two- and three-dimensional cases. 
%It is also worth noting the work \cite{Kiselev}, which is dedicated to generalized functionally invariant solutions of the wave equation.

In the present paper we seek   general solutions of equation (\ref{p}) with a variable coefficient $c^2$. Additionally, two-dimensional equations (\ref{Ptt}) are found that are equivalent to the wave equation
\begin{equation} \label{UTT} u_{tt} = u_{xx} + u_{yy}. \end{equation}
The article is structured as follows. The second section briefly describes the Laplace cascade method in the modification of Legendre and Imshenetsky \cite{Imshen, Goursat}, providing an example of applying the method to equation (\ref{p}). In the third section, a condition on the coefficient $c^2$ is derived, under which equation (\ref{p}) has a first rank solution, that is, one that can be represented in the form
$$ p = A(x) (T(I_1)+X(I_2)) + A_1(x) (T_t^{\prime}(I_1)+X_t^{\prime}(I_2))  . $$
where $T$ and $X$ are arbitrary functions of the characteristic variables
$$  I_1 = t+ \int \frac{dx}{c(x)} , \qquad   I_2 = t- \int \frac{dx}{c(x)}  ,  $$
and $A$ and $A_1$ are certain functions of $x$.
As a result, the coefficients $c^1$, with which the equation (\ref{p}), has a general solution, are obtained. 
For example, the equation
$$ p_{tt} = [(s_1x+s_2)(c_1x+c_2)^2]^{4/3}  p_{xx} ,\qquad s_1, s_2, c_1, c_2\in\mathbb{R} ,$$
generalizes two equations integrated by L. Euler.

In the fourth section, the problem of equivalence for wave equations in  two-dimensional case is considered. It is shown to which equation of the form
$$ u_{tt} = u_{xx} +s(x)u_x    $$
equation (\ref{p}) is reduced by changing the independent variable. Specific examples of variable transformations are examined. It is proven that conformal transformations map solutions of equation (\ref{Ptt}) with one coefficient to solutions of the equation with another coefficient, i.e., conformal transformations act on the set of equations of the form (\ref{Ptt}). Examples of equations reducible to the form (\ref{UTT}) are provided. All considerations are local.

\section{Elements of Laplace cascade method}

Let us consider the second-order linear hyperbolic equation with variable coefficients \begin{equation} \label{utt} u_{tt} + a_{11}u_{tx} + a_{02}u_{xx} + a_1u_{t} + a_2u_{x} + a_0u = 0, \end{equation} 
where $a_{11},a_{02},a_1,a_2,a_0$ are functions of $t,x$.
To find solutions of this equation, the Laplace cascade method, modified by Legendre and Imshenecky \cite{Imshen, Goursat}, proves useful. 

Let us briefly describe this approach. Suppose that the principal symbol of the equation 
$$ S=\xi_1^2 +a_{11}\xi_1\xi_2 + a_{02}\xi_2^2        $$
can be factored into two distinct terms as follows:
$ S=  (\xi_1+\lambda_1\xi_2) (\xi_1+\lambda_2\xi_2) . $
Then the characteristic derivatives \cite{Courant} have the following form
$$ \mathcal{D}_1 = \partial_t + \lambda_1\partial_x ,\qquad 
\mathcal{D}_2 = \partial_t + \lambda_2\partial_x .          $$
So, the equation (\ref{utt}) can be represented in two characteristic forms
$$ \mathcal{D}_1\mathcal{D}_2u +s_1 \mathcal{D}_1u+ q_1 \mathcal{D}_2u +c_1u=0,$$
$$ \mathcal{D}_2\mathcal{D}_1u +s_2 \mathcal{D}_2u+ q_2 \mathcal{D}_1u +c_2u=0,$$
where the coefficients $s_1, q_1, s_2, q_2, c_1, c_2$ are expressed through the coefficients of equation (\ref{utt}). Let us rewrite the latter equations as
\begin{equation} \label{D1D2} 
	(\mathcal{D}_1+q_1)(\mathcal{D}_2+s_1) u +(-\mathcal{D}_1(s_1)-s_1q_1+c_1)u=0,	
\end{equation}
\begin{equation} \label{D_2D_1} 
	(\mathcal{D}_2+q_2)(\mathcal{D}_1+s_2) u +(-\mathcal{D}_2(s_2)-s_2q_2+c_2)u=0 .	
\end{equation}
The functions $h= \mathcal{D}_1(s_1)+s_1q_1-c_1$, $k= \mathcal{D}_2(s_2)+s_2q_2-c_2$ 
are called Laplace invariants. 

Suppose that both Laplace invariants are zero. Then the equations (\ref{D1D2}), (\ref{D_2D_1}) factorize, and the functions $v$ and $w$ satisfy the 
first-order partial differential equations
\begin{equation} \label{FirstIn} 
	\mathcal{D}_2v+s_1 v =0, \qquad \mathcal{D}_1w+s_2 w=0 .
\end{equation}
Hence, the functions $v$, $w$, and $v+w$ are solutions to equation (\ref{utt}).
%$ and the function $u$ satisfies two first order partial derivative equations
%$$ \mathcal{D}_2u+s_1 u =0, \qquad \mathcal{D}_1u +s_2 u=0 .$$$
The equations (\ref{FirstIn}) are the intermediate integrals \cite{Goursat,Kaptsov}  for (\ref{utt}).

The solution of partial differential equations (\ref{FirstIn}) reduces to the solution of characteristic equations
\begin{equation} \label{Char} 
	\frac{dt}{1} = \frac{dx}{\lambda_1} = \frac{dv}{-s_2v}, \quad 
	\frac{dt}{1} = \frac{dx}{\lambda_2} = \frac{dw}{-s_1w}	.
\end{equation}
Note that the solutions to equations 
$$ \frac{dx}{dt} = \lambda_1,\qquad \frac{dx}{dt} = \lambda_2 $$
define the characteristics of the equation (\ref{utt}). Let us denote by $I_1(t,x)$ and $I_2(t,x)$ the integrals of the last equations. 
Then $I_1(t,x), J_1=v/A(t,x)$ and $I_2(t,x), J_2 =w/B(t,x)$ are integrals of the systems (\ref{Char}), where $A(t,x)$ , $B(t,x)$ are some functions. 
So, the general solutions of the equations (\ref{FirstIn}) have the following form
$$ v=A T(I_1) ,\qquad w=B X(I_2) , $$
where $T$ and $X$ are arbitrary functions. Hence, the solution of the equation (\ref{utt}), with zero Laplace invariants, is given by the formula
$$ u = A T + B X . $$

Let $h\neq 0$, then by performing the Laplace transform
$$ u = \frac{\mathcal{D}_1(u_1)+q_1u_1}{h}   , $$
we obtain a second order equation for the function $u_1(t,x)$. If the Laplace invariants of the new equation are not zero, we apply the Laplace transform again. Suppose that at some step both invariants are zero, then the solution to the equation (\ref{utt}) is represented as \cite{Imshen,Goursat}
\begin{equation} \label{Repr} 
	u = A T +A_1 T^{\prime} +\dots + A_n T^{(n)} + B X + B_1 X^{\prime} +\dots B_nX^{(n)} ,
\end{equation}
where $A_i, B_i$ are some particular functions of $t, x$, and $T, X$ are arbitrary functions of $I_1$ and $I_2$, respectively.

{\it Remark.} It may happen that one Laplace invariant is zero, while the other is nonzero. In that case, the representation (\ref{Repr}) should be modified; the details can be found in the literature cited above.
Following \cite{Goursat}, a solution of equation (\ref{utt}) in the form (\ref{Repr}) will be referred to as a solution of rank $n$.

As an example, consider the equation
\begin{equation} \label{uKt} 
	u_{tt} = K^2(x)u_{xx} ,
\end{equation}
where $K$ is a function of $x$, to be determined from the condition that Laplace invariants are zero.
Obviously, the the characteristic derivatives  have the form
$$ \mathcal{D}_1= \partial_t + K(x) \partial_x, \qquad \mathcal{D}_2= \partial_t - K(x)\partial_x . $$
The characteristic forms of the equation (\ref{uKt}) are as follows
$$ \mathcal{D}_1\mathcal{D}_2u +\frac{K^{\prime}}{2}(\mathcal{D}_1-\mathcal{D}_2)u=0,  
\qquad
\mathcal{D}_2\mathcal{D}_1u +\frac{K^{\prime}}{2}(\mathcal{D}_1-\mathcal{D}_2)u=0 .
$$
So the Laplace invariants are equal to
$$ h = k= \mathcal{D}_1\left(\frac{K^{\prime}}{2}\right)- 
\left(\frac{K^{\prime}}{2}\right)^2 = \frac{KK^{\prime\prime}}{2} -\left(\frac{K^{\prime}}{2}\right)^2.  $$
It is easy to see that if $h=0$, then the function $K$ has the form
$$ K = (m_1 x+m_2)^2 , $$
where  $m_1, m_2$  are arbitrary constants. 
For  $m_1\neq 0$, without loss of generality,  we can assume that $m_1=1, m_2=0$, and the corresponding equations (\ref{FirstIn}) have the form
$$ v_t +x^2v_x-xv =0 ,\qquad w_t-x^2w_x+xw=0 .$$
These equations have solutions
$$ v=xT\left(t+\frac{1}{x}\right) ,\qquad w=xX\left(t-\frac{1}{x}\right) , $$
where $T$ and $X$ are arbitrary functions.
Thus, the solution to the equation 
\begin{equation} \label{x^4} 
	u_{tt} = x^4u_{xx}	
\end{equation}
is represented as $u= x(T+X)$. 
This solution was originally obtained by Euler.

\section{First rank solutions}

In this section, we will find first-rank solutions of equation (\ref{uKt}) for certain functions
$K(x)$.  We will not compute the Laplace invariants but will directly use the representation (\ref{Repr}).

The characteristic curves of the equation (\ref{uKt}) are determined from differential equations
$$ \frac{dx}{dt} = \pm K(x) .$$
Let's introduce a new function
$$ a(x)=\int\frac{dx}{K(x)}  , $$
then the functions
\begin{equation} \label{I12} 
	I_1 = t + a(x), \qquad I_2 = t - a(x) 	
\end{equation}
are constant on the characteristics. The equation (\ref{uKt}) is rewritten as
\begin{equation} \label{ua} 
	u_{tt} = \frac{u_{xx}}{a^{\prime}(x)^2} .
\end{equation}

We will search for a solution of the equation (\ref{ua}) of rank one, i.e., represented as
\begin{equation} \label{U=} 
	u = A(x)T(I_1) +B(x)X(I_2) +A_1(x)T^{\prime}_t +B_1 X^{\prime}_t ,
\end{equation}
where $T$, $X$ are arbitrary functions of $I_1$, $I_2$ respectively, and $A, B, A_1, B_1$ some functions of $x$ to be found.
We substitute the representation (\ref{U=}) into the equation (\ref{ua}) and obtain the linear combination of the functions $T$, $X$, $T^{\prime}$, $X^{\prime}$, $T^{\prime\prime}$, $X^{\prime\prime}$,  which must be identically equal to zero.
Equating to zero the coefficients at $T$, $T^{\prime}$, $T^{\prime\prime}$,  we obtain a system of ordinary differential equations
\begin{equation} \label{T T1} 
	A^{\prime\prime} = 0,\qquad Aa^{\prime\prime} + 2A^{\prime}a^{\prime} +A_1^{\prime\prime} = 0,\qquad A_1a^{\prime\prime} +2 A^{\prime}_1a^{\prime}=0.
\end{equation}
Equating to zero the coefficients at $X$, $X^{\prime}$, $X^{\prime\prime}$, we have a second system of equations
\begin{equation} \label{X X1} 
	B^{\prime\prime} = 0, \qquad Ba^{\prime\prime} + 2B^{\prime}a^{\prime} -B_1^{\prime\prime} = 0,\qquad B_1a^{\prime\prime} +2 B^{\prime}_1a^{\prime}=0.
\end{equation}

From the first two equations of the systems (\ref{T T1}), (\ref{X X1} ) we have
\begin{equation} \label{A A1}
	A = c_1x+c_2 , \quad A_1=mx +m_1 -aA,\quad B=b_1x+b_2,\quad B_1=nx+n_1+aB ,   
\end{equation}
where $c_1, c_2, b_1, b_2, m_1, m, n_1, n$ are arbitrary constants.
It follows from the last equations of the systems  that
$$ a^{\prime} = \frac{r}{A_1^2} = \frac{q}{B_1^2} , \qquad r,q\in \mathbb{R} .    $$
We may assume without loss of generality that 
$ B=A,\ B_1=-A_1 .$
Hence, the function $a(x)$ satisfies the first order ordinary differential equation
\begin{equation} \label{a'=} 
	a^{\prime} = \frac{r}{[mx+m_1-(c_1x+c_2)a]^2} ,\qquad r \in \mathbb{R}.
\end{equation}
If we solve this equation, we get the rest of the functions by the formulas (\ref{A A1}).

Let $m=m_1=0$, then the solution of the equation (\ref{a'=}) has the form 
$$ a = \left(\frac{bc_1A-3r}{c_1A}\right)^{1/3} , $$
where $b$ is an arbitrary constant.
Hence, the coefficient of $K^2$ is written as follows
$$ K^2 = \left(\frac{bc_1A-3r}{c_1A}\right)^{4/3} \frac{A^{8/3}}{r^2} .   $$
Thus solution of the equation (\ref{ua}) is given by the formula (\ref{U=}), where
$$A = c_1x+c_2 , \quad B=A,\quad A_1=-aA,\quad B_1=-A_1 .$$
A more symmetrical notation of the previous formulas is
\begin{equation} \label{B=A}
	a = d\left(\frac{c_1x+c_2}{s_1x+s_2}\right)^{1/3},\quad K^2=[(s_1x+s_2)(c_1x+c_2)^2]^{4/3}, \quad d=\frac{3}{c_1s_2-c_2s_1} ,
\end{equation}
where $c_1, c_2, s_1, s_2$ are arbitrary constants. Thus, we have found a solution
of rank one for the equation
\begin{equation} \label{GenEuler}
	u_{tt} = [(s_1x+s_2)(c_1x+c_2)^2]^{\frac{4}{3}}u_{xx} . 
\end{equation}
Note that when $s_1=c_2=0$ , $s_2=c_1=1$ the coefficient of $K^2$ is $x^{8/3}$,
and when $s_2=c_1=0$ , $s_1=c_2=1$ is equal to $x^{4/3}$. Solutions of the equation (\ref{uKt}) for these coefficients were found by Euler \cite{Euler}.
%No other explicit solutions of the equation (\ref{a'=}) can be found.

In order to find implicit solutions of equation (\ref{a'=}), we perform a change of variables, considering $x$ as a function of $a$. As a result, we obtain the Riccati equation
\begin{equation} \label{Ric}
\frac{dx}{dy} = \frac{(mx+m_1-(c_1x+c_2)y)^2}{r} , 
\end{equation}
where $y$ denotes $a$.
Let us present several solutions of equation (\ref{Ric}) for different values of parameters.
If we take  
$$ r=1,\quad c_1= 0, \quad c_2=1,\quad m=1,\quad m_1=0 , $$ 
then the solution of equation (\ref{Ric}) will have the form
$$ x= y - \frac{be^{2y}+1}{be^{2y}-1}  ,            $$
where $b$ is an integration constant.
Assuming $b=-1$, we obtain the function
\begin{equation} \label{x=y}
	x= y - \tanh y	. 
\end{equation}
The graph of the function $x(y)$ has an oblique asymptote $x=y$. The inverse function $a(x)$ is continuous everywhere and differentiable except at the point $x=0$, where the derivative  
 $a^{\prime}$
tends to infinity. The graph of the function $a(x)$ also has an asymptote $y=x$.

If we now take
$$ r=-1,\quad c_1= 0, \quad c_2=1,\quad m=1,\quad m_1=0 , $$
then the solution of equation (\ref{Ric}) will have the form
$$ x = y - \tan(y+b) ,\qquad b\in\mathbb{R} . $$
Consequently, the function $y=a(x)$, being a solution of equation (\ref{a'=}) with some initial data, is smooth, decreasing, has two horizontal asymptotes, and its graph lies in the strip between these asymptotes. Hence, $a^{\prime}\rightarrow 0$
and $K^2(x)\rightarrow \infty $ as $|x|\rightarrow\infty$.

Here are a couple more examples of the solution of the Riccati equation (\ref{Ric}).  If
$$r=1,\ c_1=1,\ c_2=0,\ m=1,\ m_1=-1,$$
then the solution is
$$ x = \frac{b+e^{2y}}{by +(y-2)e^{2y}} , $$
where $b$ is a constant of integration and $y=a$.

If $r=c_1 =m= 1$, $c_2= m_1=-1$, then we get the solution
$$ x = -\frac{(\sqrt{3} -3y -6)S_1 +b(\sqrt{3} +3y +6)S_2}{(\sqrt{3} -3y +3)S_1 +b(\sqrt{3} +3y -3)S_2} ,\quad b\in\mathbb{R} . $$
	Here $S_1, S_2$ are given by the formulas
	$S_1 = \exp(y(2y^2+6\sqrt{3}+3y-12)/6)$, $S_2=\exp(y(-2y^2+6\sqrt{3}-3y+12)/6)$ 

It is possible to seek second-rank solutions 
\begin{equation} \label{2rang} u= A(x)(T+X) +A_1(x)(T_t -X_t) +A_2(x)(T_{tt}+X_{tt}) \end{equation} for equation (\ref{ua}).
Here, as before, the functions $T$ and $X$ are arbitrary, depending on $I_1$ and $I_2$
respectively, and the functions $A, A_1, A_2$ need to be determined.
 By substituting the representation 
(\ref{2rang}) into equation (\ref{ua}) and collecting similar terms at $T,...,T_{ttt}$,
$X,...,X_{ttt}$, we obtain a system of four ordinary differential equations 
$$	A^{\prime}=0 , \quad Aa^{\prime}=0 , \quad Aa^{\prime} +2A^{\prime}a^{\prime}+A_1^{\prime}=0 , \quad A_1a^{\prime} +2A_1^{\prime}a^{\prime}+A_2^{\prime}=0 , \quad
	A_2a^{\prime\prime} +2A_2^{\prime}a^{\prime}a^{\prime} = 0. $$
By integrating the first, second, and fourth equations of the system, we obtain 
$$ A=c_1x+c_2,\quad A_1 = mx+m_1-aA,\quad a^{\prime} = \frac{r}{A_2^2} ,\quad r\in\mathbb{R} .$$
However, the third equation remains unsimplified. Only certain particular solutions can be found.

\section{Equivalence of equations}

Two equations are equivalent if there exists an invertible change of variables that transforms the solution of one equation into the solution of the other. In general, establishing the equivalence of equations is difficult. The issues of the equivalence of differential equations are discussed in the monograph \cite{Olver}.
A classic example is the Legendre transformation \cite{Courant}
$$ \xi = u_x ,\quad \eta = u_y , \quad v= xu_x+yu_y-u , $$
which transforms solutions of the nonlinear equation
$$ a(u_x,u_y)u_{xx} +b(u_x,u_y)u_{xy} +c(u_x,u_y)u_{yy} =0 $$
into solutions of the linear equation
$$ a(\xi,\eta)v_{\eta\eta} -b(\xi,\eta)u_{\xi\eta} +c(\xi,\eta)u_{\xi\xi} =0 ,$$
if $u_{xx}u_{yy}-u_{xy}^2\neq 0$.
The inverse Legendre transformation converts solutions of a linear equation into solutions of a nonlinear equation, i.e. they are equivalent. 

%В качестве примера рассмотрим уравнение Лагранжа 

As an example, consider the Lagrange equation \cite{CF}
\begin{equation} \label{Lag} 
 x_{tt} = N^2(x_h)x_{hh} ,
\end{equation}
which describes the one-dimensional isentropic flow of a gas.
Here $N$ is the acoustic impedance, $t$ and $h$ are independent variables.
The Legendre transformation reduces the equation (\ref{Lag}) to the form
   \begin{equation} \label{LinLag} 
 	u_{tt} = K^{2}(x)u_{xx} ,
 \end{equation}
with $K=1/N$.
Therefore, knowing a solution of the equation (\ref{LinLag}), we can obtain a solution of the the equation (\ref{Lag}) and vice versa.

%Therefore, the equations
%$$ u_{yy}-c^2(u_x)u_{xx}=0 ,\qquad v_{\xi\xi} -c^2(\xi)v_{\eta\eta} =0.$$
%are equivalent. 

Let us once again consider equation (\ref{ua}). It is easy to see that a change of the independent variable $y=a(x)$
 transforms equation (\ref{ua}) into the form 
 \begin{equation} \label{vtt=} 
 	v_{tt}= v_{yy} +\frac{a^{\prime\prime}}{(a^{\prime})^2}v_y . 
 \end{equation} 
If the derivative of $a^{\prime}(x)$ is not zero, then the inverse function $x=a^{-1}(y)$ exists and we have
$$ a^{\prime} = \frac{1}{x^{\prime}} \qquad
a^{\prime\prime} = -\frac{x^{\prime\prime}}{(x^{\prime})^3} . $$
So the equation (\ref{vtt=}) is written in the form
$$ v_{tt}= v_{yy} -\frac{x^{\prime\prime}}{x^{\prime}}v_y . $$

Here are some examples of transformations. Equation
$$ u_{tt} = x^{\alpha}u_{xx} \qquad \alpha\in\mathbb{R} , $$
by substituting $$y=\frac{x^{1-\alpha}}{1-\alpha}  $$
%$$y=\frac{x^{x{1-\alpha}}{1-\alpha} $$ 
	is reduced to the Euler-Poisson-Darboux equation.
	$$ v_{tt} = v_{yy} + \frac{\alpha}{\alpha -1}v_y . $$
	For special values of $\alpha$, Euler \cite{Euler} found solutions of arbitrary rank of this equation.  Applications of these solutions are known in gas dynamics \cite{Mises,Lan}. Recently, other applications have appeared \cite{Pel1,Pel2}.
	
As a second example, consider equation (\ref{GenEuler})
$$ u_{tt} = [(s_1x+s_2)(c_1x+c_2)^2]^{\frac{4}{3}}u_{xx} . $$ 
The solution to this equation was obtained in the previous paragraph. Substitution
$$ y = q\left(\frac{c_1x+c_2}{s_1x+s_2}\right)^{1/3} $$
reduces this equation to the form
$$ v_{tt} =v_{yy} -2\frac{c_1q^3+2s_1y^3}{y(c_1q^3-s_1y^3)}v_y , $$
where $q=3/(c_1s_2-c_2s_1)$. So, the last equation also has a solution of rank one.

Consider another previously obtained equation
\begin{equation} \label{tanh4}
	u_{tt}= \tanh^4(y)u_{xx}	,
\end{equation}
where $x$ and $y$ are related by the transformation 
$x =y - \tanh y$. This transformation reduces equation (\ref{tanh4}) to the form
$$ v_{tt} =v_{yy} - \frac{4}{\sinh(2y)}v_y .$$
In the article \cite{KaptsovPMM} a solution to the equation 
$$ v_{tt} =v_{yy} - \frac{2n}{\sinh(y)}v_y $$
was found for any integer $n$.

Let's now consider two-dimensional equations
\begin{equation} \label{u2D}
	u_{tt}= c^2(x,y) (u_{xx}+u_{yy})	.
\end{equation}
An interesting question is when this equation can be transformed into a similar equation
\begin{equation} \label{v2D}
	v_{tt}= c_1^2(x,y) (v_{x_{1}x_1}+u_{y_1y_1})	.
\end{equation}
using the change of variables
\begin{equation} \label{trans}
	x_1 = f(x,y) ,\qquad y_1=g(x,y) ,\qquad v(t,x_1,y_1)=u(t,x,y) .
\end{equation}
First we find the second derivatives
$$ u_{tt}=v_{tt} ,\quad u_{xx}=f_x^2v_{x_1x_1} +2f_xg_xv_{x_1y_1} +g_x^2v_{y_1y_1} +f_{xx}v_{x_1} +g_{xx}v_{y_1} ,  $$
$$ u_{yy} =f_y^2v_{x_1x_1} +2f_yg_yv_{x_1y_1} +g_y^2v_{y_1y_1} +f_{yy}v_{x_1} +g_{yy}v_{y_1} . $$
Substituting these expressions into (\ref{u2D}), we obtain equation
$$ v_{tt} =c^2[(f_x^2+f_y^2)v_{x_1x_1} + (g_x^2+g_y^2)v_{y_1y_1} +2(f_xg_x+f_yg_y)v_{x_1y_1} +(f_{xx}+f_{yy})v_{x_1} + (g_{xx}+g_{yyy})v_{y_1}]
$$
For this equation to coincide with (\ref{v2D}), the following conditions must be satisfied
\begin{equation} \label{Eqfg}
	f_x^2+f_y^2 = g_x^2+g_y^2 ,\quad f_xg_x+f_yg_y=0,\quad
	f_{xx}+f_{yy}=0, \quad g_{xx}+g_{yy}=0	,
\end{equation}
\begin{equation} \label{c=c1}
	c^2 (f_x^2+f_y^2) =c_1^2.
\end{equation}

Solving the first two equations of the system (\ref{Eqfg}), with respect to $f_x, f_y$, we obtain
\begin{equation} \label{CR}
	f_x^2=\pm g_y , \quad f_y=\mp g_x .
\end{equation}
This means that the pair of functions $f, g$ defines an conformal mapping in the Euclidean space $\mathbb{R}^2(x,y)$. 
Obviously, the third and fourth equations of system (\ref{Eqfg}) hold due to the relations (\ref{CR}).

Of particular interest are equations of the form (\ref{u2D}), which are reduced by a comfortable transformation to an equation with a constant coefficient $c_1=1$. Since the inverse transformation to a conformal transformation is also conformal, it is sufficient to indicate the equations that are obtained from the equation
\begin{equation} \label{wave}
	u_{tt} = u_{xx} + u_{yy} .
\end{equation}

As is well known \cite{Remmert}, any holomorphic function $F(z)$ defines conjugate harmonic functions $f$ and $g$ that satisfy the Cauchy-Riemann conditions, i.e., it defines a conformal mapping of the first kind. Similarly, any analytic function $F(\bar{z})$ of the variable $\bar{z}$ defines a antiholomorphic function. 

Take as an example the antiholomorphic function $F(\bar{z})=1/\bar{z}$. Then its real and imaginary parts have the form
$$ f=\frac{x}{x^2+y^2} ,\quad g=\frac{y}{x^2+y^2} . $$
This mapping is an inversion with respect to a circle of unit radius. The inverse mapping is given by similar formulas
$$ x=\frac{x_1}{x_1^2+y_1^2} ,\quad y=\frac{y_1}{x_1^2+y_1^2} . $$
If the coefficient $c^2$ in equation (\ref{u2D}) is equal to one, then according to (\ref{c=c1}), we find
$$ c_1^2 = f_x^2+f_y^2 = \frac{1}{(x^2+y^2)^2} = (x_1^2+y_1^2)^2 . $$
Therefore, the corresponding equation (\ref{v2D}) has the form
\begin{equation} \label{x^2}
	v_{tt} =(x_1^2+y_1^2)^2(v_{x_1x_1}+v_{y_1y_1}) .
\end{equation}
This is a two-dimensional analogue of equation (\ref{x^4}).

Now let the holomorphic function $F$ be equal to
$$ F(z) = e^z = e^x\cos y + ie^x\sin y . $$
From here, we obtain the change of variables
\begin{equation} \label{Exp}
	x_1 = e^x\cos y , \quad y_1=e^x\sin y .
\end{equation}
and the coefficient
$$c_1^2 = f_x^2+f_y^2=e^{2x}= x_1^2+y_1^2 .$$
So, in this case, the equation (\ref{v2D}) is as follows
\begin{equation} \label{vx}
	v_{tt} =(x_1^2+y_1^2)(v_{x_1x_1}+v_{y_1y_1}) .
\end{equation}
Therefore, knowing the solution of the wave equation (\ref{wave}), it is easy to obtain solutions of the equations (\ref{x^2}) and (\ref{vx}). The number of such examples can easily be increased.
Among the many particular solutions of the equation (\ref{wave}), we will single out the well-known Smirnov-Sobolev solution \cite{Sobolev}.

Let us briefly consider the three-dimensional equation
\begin{equation} \label{3Dwave}
	u_{tt} =c^2(x,y,z)( u_{xx} + u_{yy} + u_{zz}) . 
\end{equation}
 If we look for a map
 $$ x_1=f(x,y,z) ,\quad y_1=g(x,y,z) ,\quad z_1=h(x,y,z) ,\quad v=\alpha(x,y,z) u , $$
 transforming the equation (\ref{3Dwave}) into the equation
 $$ v_{tt} =c_1^2(x_1,y_1,z_1)( v_{x_1x_1} + v_{y_1y_1} + v_{z_1z_1}) , $$
 then we can see that such transformations are related to the ten-dimensional symmetry group of the three-dimensional Laplace equation. 
    The point is that the expression
$$ u_{xx} + u_{yy} + u_{zz} $$
under the action of this group of symmetries transforms into an expression of the form
$$ ( v_{x_1x_1} + v_{y_1y_1} + v_{z_1z_1})b $$ 
with some function $b(x_1,y_1,z_1)$. This conformal group is a subgroup of the equivalence group of equation (\ref{3Dwave}). 

 In particular, the Laplace equation is invariant under the transformation
$$ x_1 = \frac{x}{r^2} \quad y_1 = \frac{y}{r^2} \quad z_1 = \frac{z}{r^2} \quad
v = \frac{u}{r} ,$$
with $r^2 = x^2+y^2+z^2$.  By means of   this transformation solutions of the 
equation  
$$ u_{tt} = u_{xx} + u_{yy} + u_{zz} . $$
are converted to solutions of the equation
$$ v_{tt} =(x_1^2+y_1^2+z_1^2)^2( v_{x_1x_1} + v_{y_1y_1} + v_{z_1z_1}) . $$
The latter equation is a three-dimensional version of the two-dimensional equation (\ref{x^2}). 
Note that in the two-dimensional case the group of conformal transformations is infinite dimensional.
It is a subgroup of the equivalence group of the equation (\ref{u2D}).

\section{Conclusion} 
 
This paper presents a method for solving linear partial differential equations in two independent variables. This approach utilizes a first-order solution representation within the Laplace cascade method. As a demonstration, we apply this method to the acoustic wave equation in a non-homogeneous medium. However, the applicability of this method extends beyond this specific equation, and it could be profitably applied to the equations explored in \cite{Churilov,Stepan}.
A different approach is founded on equivalence transformations of linear equations. This method is applied to two- and three-dimensional acoustic equations. Examples of equations equivalent to the constant coefficient wave equation are derived.

In conclusion, a few words about spherical and cylindrical waves in inhomogeneous medium. Propagation of spherical wave  is described by the equation
\begin{equation} \label{sphera}
	p_{tt}= c^2(r)\left(p_{rr}+ \frac{2}{r}p_r\right ).	
\end{equation}
Introducing a new function $v=rp$ we obtain the equation
$$ v_{tt}= c^2(r)v_{rr} .   $$
If we consider the equation for cylindrical waves
\begin{equation} \label{cylindr}
	p_{tt}= c^2(r)\left(p_{rr}+ \frac{1}{r}p_r\right ).	
\end{equation}
then the replacement of the variable $r=\exp(y)$ leads this equation to the form
$$ v_{tt}= \frac{c^2(e^y)}{e^{2y}}v_{yy} .   $$

Hence, all solutions found above for equation (\ref{uKt}) can also be applied to equations   (\ref{sphera}) and (\ref{cylindr})

\end{document}